# Interplay between Velocity and Travel Distance of Kinesin-based Transport in the Presence of Tau

Jing Xu,* Stephen J King,† Maryse Lapierre-Landry,* Brian Nemec*

*Physics Graduate Group, University of California, Merced, California, USA; †Burnett School of Biomedical Sciences, University of Central Florida, Florida, USA

ABSTRACT   Although the disease-relevant microtubule-associated protein tau is known to severely inhibit kinesin-based transport *in vitro*, potential mechanisms for reversing this detrimental effect to maintain healthy transport in cells remain unknown. Here we report the unambiguous up-regulation of multiple-kinesin travel distance despite the presence of tau, via decreased single-kinesin velocity. Interestingly, the presence of tau also modestly reduced cargo velocity in multiple-kinesin transport, and our stochastic simulations indicate that the tau-mediated reduction in single-kinesin travel underlies this observation. Taken together, our observations highlight a non-trivial interplay between velocity and travel distance for kinesin transport, and suggest that single-kinesin velocity is a promising experimental handle for tuning the effect of tau on multiple-kinesin travel distance.

Conventional kinesin is a major microtubule-based molecular motor that enables long-range transport in living cells. Although traditionally investigated in the context of single-motor experiments, two or more kinesin motors are often linked together to transport the same cargo *in vivo* (1-4). Understanding the control and regulation of the group function of multiple kinesins has important implications for reversing failure modes of transport in a variety of human diseases, particularly neurodegenerative diseases. Tau is a disease-relevant protein enriched in neurons (5-6). The decoration of microtubules with tau is known to strongly inhibit kinesin transport *in vitro* (7-9), but how kinesin-based transport is maintained in the presence of high levels of tau, particularly in healthy neurons, remains an important open question. To date, no mechanism has been directly demonstrated to reverse the inhibitory effect of tau on kinesin-based transport. Here we present a simple *in vitro* study that demonstrates the significant up-regulation of multiple-kinesin travel distance with decreasing ATP concentration, despite the presence of tau.

The current investigation was motivated by our recent finding that single-kinesin velocity is a key controller for multiple-kinesin travel distance along bare microtubules (10). The active stepping of each kinesin motor is stimulated by ATP (11), and each kinesin motor remains strongly bound to the microtubule between successive steps (10-11). As demonstrated for bare microtubules (10), with decreasing ATP concentrations, each microtubule-bound kinesin experiences a decreased stepping rate per unit time and spends an increased fraction of time in the strongly bound state; additional unbound kinesins on the same cargo have more time to bind to the microtubule before cargo travel terminates. Thus, reductions in single-kinesin velocity increase the probability that at least one kinesin motor will remain bound to the microtubule per unit time, thereby increasing the travel distance of each cargo (10). Since this effect only pertains to the stepping rate of each individual kinesin and does not address the potential presence of roadblocks such as tau on the microtubules, we hypothesized in the current study that single-kinesin velocity may be exploited to relieve the impact of tau on multiple-kinesin travel distance.

We focused our *in vitro* investigation on human tau 23 (htau23, or 3RS tau), an isoform of tau that exhibits the strongest inhibitory effect on kinesin-based transport (7-9). Importantly, htau23 does not alter the stepping rate of individual kinesins (7, 9), supporting our hypothesis and enabling us to decouple single-kinesin velocity from the potential effects of tau. We carried out multiple-kinesin motility experiments using polystyrene beads as *in vitro* cargos (8, 10), ATP concentration as an *in vitro* handle to controllably tune single-kinesin velocity (10-11), and three input kinesin concentrations to test the generality of potential findings for multiple-kinesin transport. Combined with previous two-kinesin studies (10, 12), our measurements of travel distance (Fig. 1A) indicate that the lowest kinesin concentration employed (0.8 nM) corresponds to an average of ~2-3 kinesins per cargo. Note that in the absence of tau, the observed decrease in bead velocity at the higher kinesin concentrations (Fig. 1A) is consistent with a recent *in vitro* finding (13). At 1 mM ATP, htau23 reduced kinesin-based travel distance by a factor of two or more (Fig. 1A&B). This observation is in good agreement with previous reports (7-8).

Consistent with our hypothesis, reducing the available ATP concentration to 20 µM increased the multiple-kinesin travel distance by more than 1.4-fold for all three input kinesin concentrations (Fig. 1B&C), despite the presence of htau23. The corresponding reduction in single-kinesin velocity with decreasing ATP concentration (10-11) is reflected in the ~3.4-fold reduction in the measured bead velocities (Fig. 1B&C). Therefore, the strong negative relationship between single-kinesin velocity and multiple-kinesin travel distance occurs not only for bare microtubules (10), but also for tau-decorated microtubules.

What causes the observed increase in travel distance at the lower ATP concentration (Fig. 1B&C)? In addition to the mechanism discussed above for the case of bare microtubules (10), an intriguing mechanism was suggested by recent studies of tau-microtubule interactions in which htau23 was observed to dynamically diffuse along microtubule lattices (14-15): reducing the stepping rate of a microtubule-bound kinesin may effectively increase the probability that a tau roadblock can diffuse away before the kinesin takes its next step.

Perhaps surprisingly, although htau23 does not impact single-kinesin velocity (7, 9), we observed a modest reduction in the average velocity of multiple-kinesin transport in experiments using tau-decorated microtubules (Fig. 1A&B). This decreased velocity reflects a substantially larger variance in the instantaneous velocity for bead trajectories in the presence of htau23 (Fig. S1 in the Supporting Material), as quantified by parsing each bead trajectory into a series of constant-velocity segments using a previously developed automatic software incorporating Bayesian statistics (16).

To test the possibility that single-kinesin travel distance impacts multiple-kinesin velocity, we performed stochastic simulations (see Supporting Material) that assumed $N$ identical kinesin motors available for transport and included kinesin's detachment kinetics (17). Previously, this model successfully captured multiple-dynein travel distances *in vivo* using single-dynein characteristics measured *in vitro* (18). In the current study, we introduced one (and only one) free parameter to reflect the probability of each bound kinesin encountering tau at each step. When encountering tau, each kinesin has a 54% probability of detaching from the microtubule (interpolated from Figure 2A of reference (7)); the undetached kinesin is assumed to remain engaged in transport and completes its step along the microtubule despite the presence of tau.

Remarkably, our simple simulation suggested that the tau-mediated reduction in single-kinesin travel is sufficient to reduce multiple-kinesin velocity (Fig. 2A). The majority of the velocity decrease is predicted to occur at the transition from single-kinesin to two-kinesin transport (Fig. 2). Further decreases in cargo velocity with increasing motor number are predicted to be modest and largely independent of tau (Fig. 2B). The results of our simulation remain qualitatively the same when evaluated at two bounds (40% and 65%) encompassing the interpolated 54% probability of kinesin detaching at tau (Fig. S2).

We note that our simple simulations do not consider the possibility that kinesin may pause in front of a tau roadblock, as previously reported (7). We omitted this consideration because the interaction strength between kinesin and the microtubule in such a paused state is unknown. In a multiple-motor geometry, could a paused kinesin be dragged along by the other motors bound to the same cargo? Could a tau roadblock be forcefully "swept" off the microtubule surface by the collective motion of the cargo-motor complex? Significant experimental innovations are necessary to specifically address these questions in future multiple-motor assays and to guide modeling efforts. Nonetheless, our simple simulation demonstrates that reducing single-kinesin travel distance is sufficient to decrease multiple-kinesin travel distance.

Taken together, our observations highlight a non-trivial interplay between velocity and travel distance for kinesin-based transport in the presence of tau. We uncover a previously unexplored dual inhibition of tau on kinesin-transport: in addition to limiting cargo travel distance, the tau-mediated reduction in single-kinesin travel distance also leads to a modest reduction in multiple-kinesin velocity. We provide the first demonstration of the unambiguous up-regulation of multiple-kinesin travel distance despite the presence of tau, via reducing single-kinesin velocity, suggesting a mechanism that could be harnessed for future therapeutic interventions in diseases that result from aberrant kinesin-based transport.

**FIGURE LEGENDS**

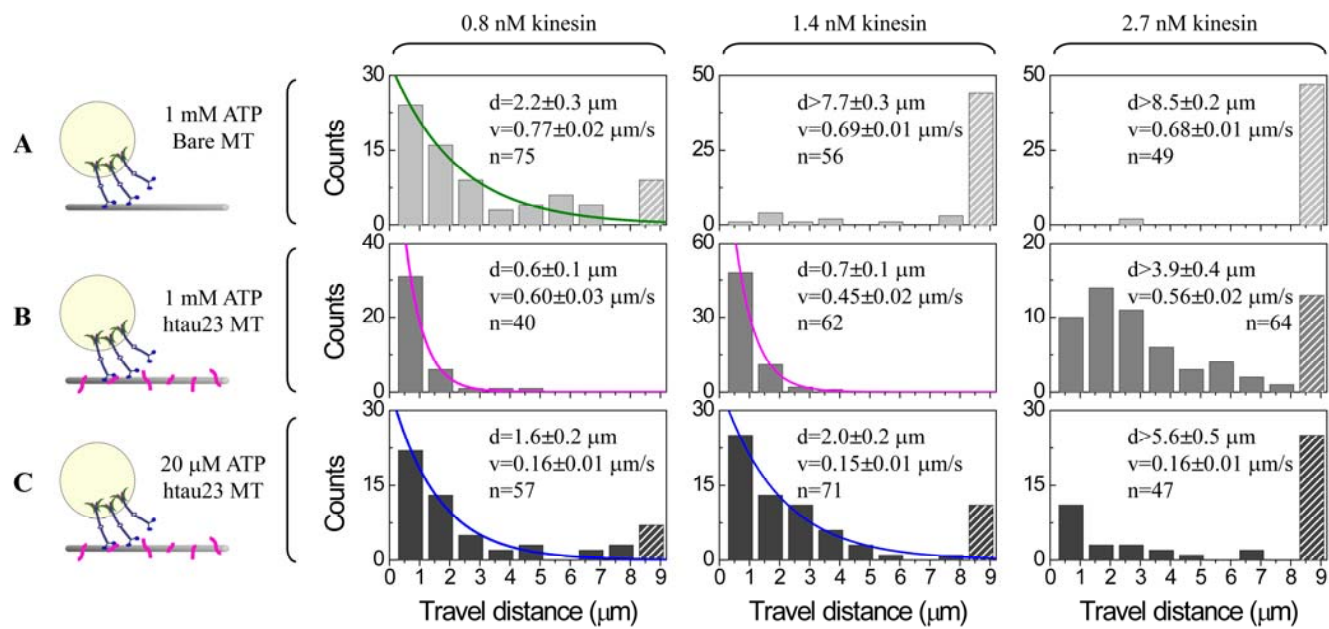

**Figure 1.** Distributions of multiple-kinesin travel distances measured at three experimental conditions, to verify the effect of tau (A&B) and to investigate the impact of single-kinesin velocity on the tau effect (B&C). Shaded bars at 8.7 μm indicate counts of travel exceeding the field of view. The mean travel distance (d; ±standard error of mean, SEM), sample size (n), and corresponding mean velocity (v; ±SEM) are indicated. MT, microtubule. Mean travel distance increased substantially at 20 μM ATP (C), despite the presence of htau23. This effect persisted across all three kinesin concentrations tested (left to right).

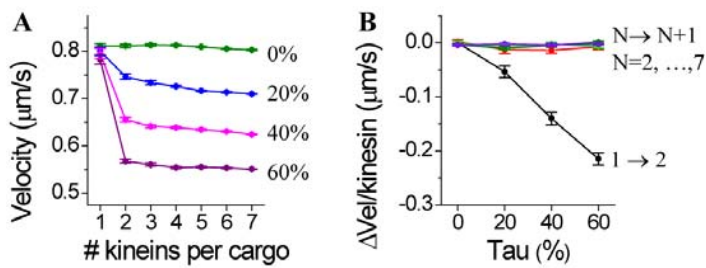

**Figure 2.** Stochastic simulations predict a tau-dependent reduction in multiple-kinesin velocity, assuming that the only effect of tau is to prematurely detach kinesin from the microtubule (or, to reduce single-kinesin travel distance). (A) Average velocity of cargos carried by the indicated number of kinesins was evaluated at 1 mM ATP and for four probabilities that a kinesin may encounter tau at each step. Mean velocity (error bars, SEM) was evaluated using 600 simulated trajectories for all simulation conditions. (B) Change in cargo velocity with each additional kinesin (ΔVel/kinesin) as a function of tau encounter probability. These values were calculated from cargo velocities shown in panel (A). Error bars indicate SEM.

**SUPPORTING MATERIAL**

Materials and Methods, and two figures are available online.

**ACKNOWLEDGEMENTS**

We thank Dr. Christopher L. Berger for helpful discussions. This work was supported in part by NIH grant NS048501 to SJK.

# Interplay between Velocity and Travel Distance of Kinesin-based Transport in the Presence of Tau

Jing Xu,*§ Stephen J King,† Maryse Lapierre-Landry,* Brian Nemec*

*Department of Physics, School of Natural Sciences, University of California, Merced, California, USA; †Burnett School of Biomedical Sciences, College of Medicine, University of Central Florida, Florida, USA

§Correspondence: jing.xu@ucmerced.edu

**SUPPORTING MATERIAL**

**MATERIALS AND METHODS**

*In Vitro* **Motility Assay**

Recombinant human tau 23 (htau23), bovine brain tubulin, and bovine brain conventional kinesin were purified as previously described (1). Htau23 was flash frozen in 1x PM buffer (100 mM PIPES, 1 mM $MgSO_4$, 2 mM EGTA, pH 6.9). Kinesin was flash frozen in 1x PMEE buffer (35 mM PIPES, 5 mM $MgSO_4$, 1 mM EGTA, 0.5 mM EDTA, pH 6.6) supplemented with 45% glycerol and 1 mM DTT, without ATP. Chemicals were purchased from Sigma.

Taxol-stabilized microtubules were polymerized by incubating 12 μM tubulin in 1x PM buffer (supplemented with 22 μM taxol and 1.1 mM GTP) for 20 min at 37 °C, diluted to 960 nM in PMEE buffer (supplemented with 20 μM taxol and 1 mM GTP), then incubated for 20 min at 37 °C with an equal volume of PMEE buffer (pH 7.1, supplemented with 20 μM taxol and 1 mM GTP) containing 0 nM or 368 nM htau23. The prepared microtubules (bare or decorated with htau23) were anchored to coverslip surfaces preincubated with poly-L-lysine as previously described (1-2). This concentration of microtubules was empirically determined to maintain sufficient numbers of isolated and parallel microtubules (~3-5 microtubules) extending across our field of view (20 μm × 20 μm, imaged at 100× magnification) (data not shown). The concentration of tau used here corresponds to a 1:20 ratio of bound tau to tubulin dimer, calculated using a previously developed model and binding parameters for htau23 (3). The resulting level of tau decoration of the microtubule is within the physiological level of tau decoration in cells (3-4), which has been demonstrated to strongly inhibit kinesin-based transport *in vitro* (1, 5); this inhibition was verified in the current study (Fig. 1A&B).

Purified kinesin was diluted to 0.8 nM, 1.4 nM, and 2.7 nM in motility buffer (66.4 mM PIPES, 50 mM $CH_3CO_2K$, 3.4 mM $MgSO_4$, 0.8 mM DTT, 0.84 mM EGTA, 10.1 μM taxol, pH 6.9) and incubated with carboxylated polystyrene beads (Polysciences) at a fixed concentration of $7.1 \times 10^5$ beads/μL for 10 min at room temperature. This bead concentration was empirically determined to optimize the number of beads in our field of view for optical trapping (data not shown). We observed that the use of 0.2 μm-diameter beads facilitated the clustering of multiple kinesin motors more effectively than the larger 0.5-1 μm-diameter beads typically used in optical-trapping experiments (data not shown), likely due to the reduced surface area of the smaller beads available for kinesin recruitment. Although this reduced bead size limits our ability to use force measurements to determine the number of motors per cargo, the lowest number of kinesins employed (0.8 nM) corresponds to ~2-3 kinesins per bead, based on previous studies of two-kinesin travel distance (6-7). Immediately following incubation, the kinesin/bead mixture was supplemented with an oxygen-scavenging solution (250 μg/mL glucose oxidase, 30 μg/mL catalase, 4.6 mg/mL glucose) and ATP (20 μM or 1 mM as indicated) prior to motility experiments.

A single-beam optical trap was used to facilitate motility measurements. An optical trap similar to that previously described (2) was custom-constructed in the Xu lab and is integrated with differential interference contrast microscopy. The optical trap (~20 mW) was used to position individual kinesin-coated beads in the vicinity of a microtubule for 30 s. During this wait time, if the bead exhibited directed motion along the microtubule, the optical trap was manually shut off to allow motility measurements in the absence of load. Bead trajectories were imaged at 100× magnification and recorded using a Giga-E camera (Basler SCA640-70GM) at 30 Hz.

Three parallel experimental conditions were employed: bare microtubules with 1 mM ATP, tau-decorated microtubules with 1 mM ATP, and tau-decorated microtubules with 20 µM ATP. Identical motor/bead mixes were used in parallel experiments, and each experimental condition was repeated a minimum of three times; the measurement order of each experimental condition was shuffled to eliminate potential artifacts.

**Data Analysis**

Bead trajectories were particle-tracked to 10-nm resolution (1/3 pixel) using template matching as previously described (1, 8). For each experimental condition, the average bead travel distance was determined by fitting a single exponential decay to the measured travel distribution where appropriate, or by the minimum algorithmic mean and associated error for travel distributions in which a significant population of beads escaped out of the field of view (20 µm × 20 µm). To account for the time that had lapsed during the manual shutoff of the optical trap, only bead trajectories with more than 250 nm of motion were analyzed (≥40 bead trajectories for each experimental condition).

In the presence of htau23, bead trajectories exhibited substantially larger variance in their instantaneous velocities. To quantify instantaneous velocity without operator bias, we parsed each bead trajectory into a series of constant velocity segments, using a previously developed automatic software incorporating Bayseian statistics (9). Only portions of each bead trajectory greater than 250 nm were used to evaluate velocity under no load. The velocity for each bead trajectory was calculated using the algorithmic mean of the parsed velocities. The average velocity for each experimental condition was determined by the algorithmic mean and the associated error for the velocity of all associated measured trajectories.

**Simulation of Multiple-kinesin Transport**

A previously developed Monte Carlo simulation model (10) was adapted to evaluate the effect of tau on single- and multiple-kinesin velocity at 1 mM ATP. Previously, this model successfully captured multiple-dynein travel distances *in vivo* using single-dynein characteristics measured *in vitro* (8). Briefly, in this model, $N$ identical motors are bound to the same cargo for transport. A travel trajectory is initiated when at least one of these $N$ motors becomes stochastically bound to the microtubule (characterized by the motor's on-rate), and is terminated when all $N$ motors become detached (characterized by an off-rate). To mimic the unloaded travel condition utilized in our experiments, no external opposing force is assumed in the model. Intra-motor strain is reflected in the load-dependence of each individual kinesin's off-rate and velocity. We utilized an unloaded stepping rate of 800 nm/s for each kinesin, and employed the same motor stiffness (0.32 pN/nm), on-rate estimation (5/s), step size (8 nm), unloaded off-rate (1/s), and formulas for off-rate and velocity under load as previously described (10).

The only free parameter in our simulation is the probability of each kinesin motor encountering a tau roadblock at each step, which we set to be 0%, 20%, 40%, or 60%. When each bound kinesin encounters a tau roadblock, it is assumed to have a 54% probability of detaching from the microtubule based on a previous single-kinesin study (interpolated from Figure 2A of reference (5)); the undetached kinesin is assumed to remain engaged in transport and completes its step along the microtubule despite the presence of tau. For simulations in Figure S2, we set the probability of a kinesin detaching from the microtubule at a tau roadblock to be 40%, 54%, and 65%.

Simulated travel trajectories longer than 180 nm and 0.18 s were analyzed. Each simulation combination resulted in 600 trajectories for analysis. The velocity of each simulated trajectory was determined using a linear fit to the corresponding bead position *vs.* time information. The average velocity for each simulation condition was determined by the algorithmic mean and associated error for the velocity of all associated simulated trajectories.

**SUPPORTING FIGURES**

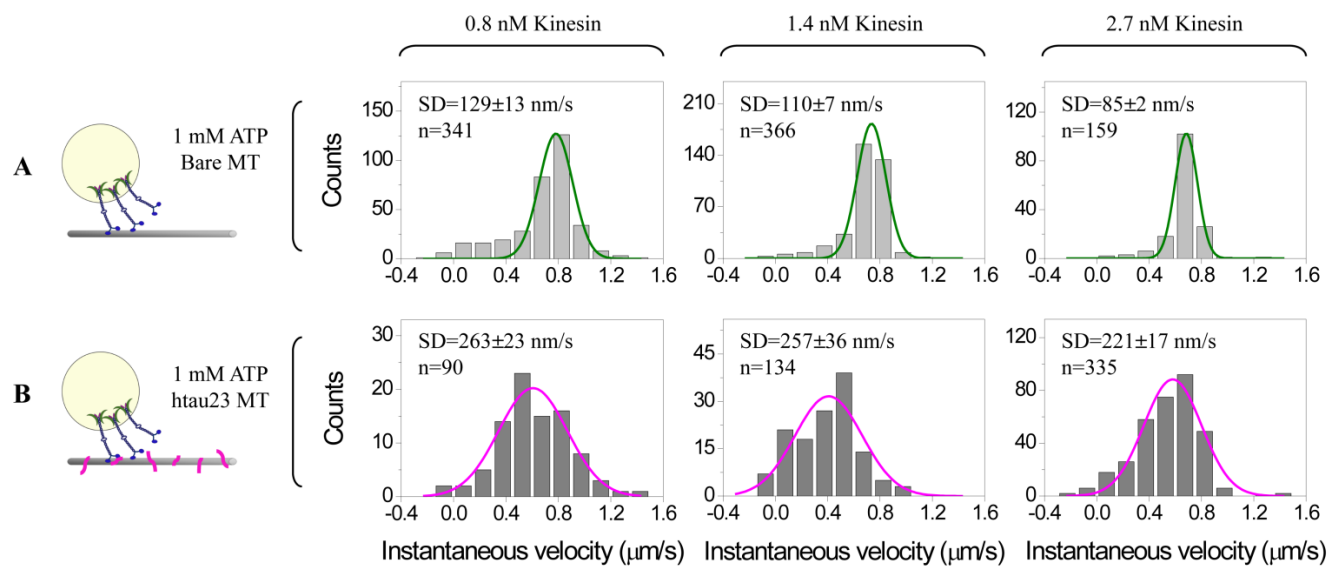

**Figure S1. The presence of htau23 substantially broadens the distribution of instantaneous velocities in multiple-kinesin transport.** Distributions of parsed instantaneous velocities for bead trajectories along bare (A) and htau23-decorated microtubules (B), measured at 1 mM ATP and corresponding to Fig. 1A&B, are each fitted to a Gaussian distribution. Indicated are the fitted standard deviation (SD, ± fitting uncertainty) and n values. MT, microtubules. We interpret the shift toward lower-velocity values in experiments using htau23 (B) as temporary backward repositioning of the cargo caused by premature detachment of bound kinesins (referred to as "bead flop" in a previous study (11)) when encountering tau. Although some reversal motion was observed in these distributions (counts below 0 μm/s), we calculate that ≤5% of the measured traces exhibited reversal motion (≥50 nm in travel duration) for each experimental condition. This reversal fraction appears to not require the presence of htau23, and is substantially lower than that reported in (11) (10-50% depending on crowding conditions), where active kinesin motors served as crowding factors on the microtubules.

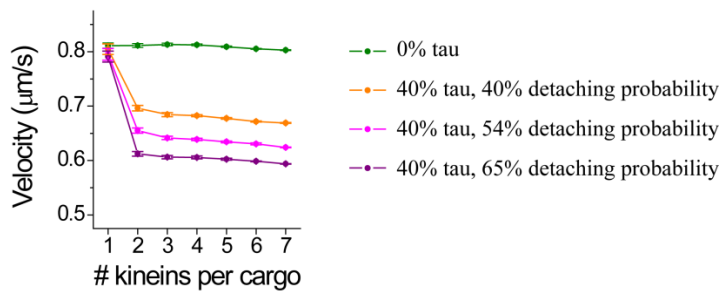

**Figure S2. The results of our simulation remain qualitatively the same when evaluated at three probabilities of a kinesin detaching at tau: 40%, 54%, and 65%.** Average velocity of cargos carried by the indicated number of kinesins was evaluated at 1 mM ATP and for two probabilities that a bound kinesin may encounter tau at each step (0% and 40% tau). Mean velocity was evaluated using 600 simulated trajectories for all simulation conditions. Error bars indicate SEM.